\pdfoutput=1
\documentclass[%
 aip,
 amsmath,amssymb,
 reprint,
]{revtex4-2}

\usepackage[T1]{fontenc}
\usepackage[utf8]{inputenc}
\usepackage{mathptmx}
\usepackage{dcolumn}
\usepackage{bm}
\DeclareUnicodeCharacter{FF5E}{\textasciitilde}
\usepackage{xcolor}
\usepackage{graphicx}
\usepackage{siunitx} 
\usepackage{braket}

\begin{document}

\preprint{AIP/123-QED}

\title[
Microwave Output Stabilization of a Qubit Controller via Device-Level Temperature Control
]{
Microwave Output Stabilization of a Qubit Controller via Device-Level Temperature Control
}

\author{Yoshinori~Kurimoto}
\email{kurimoto@quel-inc.com}
\affiliation{
QuEL, Inc., Hachioji ON Building 5F, 4-7-14 Myojincho, Hachioji, Tokyo, Japan
}

\author{Dongjun~Lee}
\affiliation{
QuEL, Inc., Hachioji ON Building 5F, 4-7-14 Myojincho, Hachioji, Tokyo, Japan
}

\author{Koichiro~Ban}
\affiliation{
QuEL, Inc., Hachioji ON Building 5F, 4-7-14 Myojincho, Hachioji, Tokyo, Japan
}

\author{Shinichi~Morisaka}
\affiliation{
QuEL, Inc., Hachioji ON Building 5F, 4-7-14 Myojincho, Hachioji, Tokyo, Japan
}
\affiliation{
Center for Quantum Information and Quantum Biology, The University of Osaka, 1-2 Machikaneyama, Toyonaka, Osaka, Japan
}

\author{Toshi~Sumida}
\affiliation{
QuEL, Inc., Hachioji ON Building 5F, 4-7-14 Myojincho, Hachioji, Tokyo, Japan
}

\author{Hidehisa~Shiomi}
\affiliation{
QuEL, Inc., Hachioji ON Building 5F, 4-7-14 Myojincho, Hachioji, Tokyo, Japan
}
\affiliation{
Center for Quantum Information and Quantum Biology, The University of Osaka, 1-2 Machikaneyama, Toyonaka, Osaka, Japan
}

\author{Yosuke~Ito}
\affiliation{
QuEL, Inc., Hachioji ON Building 5F, 4-7-14 Myojincho, Hachioji, Tokyo, Japan
}

\author{Yuuya~Sugita}
\affiliation{
QuEL, Inc., Hachioji ON Building 5F, 4-7-14 Myojincho, Hachioji, Tokyo, Japan
}

\author{Makoto~Negoro}
\affiliation{
QuEL, Inc., Hachioji ON Building 5F, 4-7-14 Myojincho, Hachioji, Tokyo, Japan
}
\affiliation{
Center for Quantum Information and Quantum Biology, The University of Osaka, 1-2 Machikaneyama, Toyonaka, Osaka, Japan
}

\author{Ryutaro~Ohira}
\email{ohira@quel-inc.com}
\affiliation{
QuEL, Inc., Hachioji ON Building 5F, 4-7-14 Myojincho, Hachioji, Tokyo, Japan
}

\author{Takefumi~Miyoshi}
\affiliation{
QuEL, Inc., Hachioji ON Building 5F, 4-7-14 Myojincho, Hachioji, Tokyo, Japan
}
\affiliation{
Center for Quantum Information and Quantum Biology, The University of Osaka, 1-2 Machikaneyama, Toyonaka, Osaka, Japan
}
\affiliation{
e-trees.Japan, Inc. Daiwaunyu Building 2F, 2-9-2, Owada-machi, Hachioji, Tokyo, Japan
}

\date{\today}

\begin{abstract}
We present the design and performance of QuEL-1 SE, which is a multichannel qubit controller developed for superconducting qubits. 
The system incorporates the active thermal stabilization of critical analog integrated circuits, such as phase-locked loops, amplifiers, and mixers, to suppress the long-term amplitude and phase drift. 
To evaluate the amplitude and phase stability, we simultaneously monitor 15 microwave output channels over 24 h using a common analog-to-digital converter.
Across the channels, the normalized amplitude exhibits standard deviations of 0.09\%--0.22\% (mean: 0.15\%), and the phase deviations are 0.35$^\circ$--0.44$^\circ$ (mean: 0.39$^\circ$).
We further assess the impact of these deviations on quantum gate operations by estimating the average fidelity of an $X_{\pi/2}$ gate under the coherent errors corresponding to the deviations.
The resulting estimated contribution to the $X_{\pi/2}$ gate infidelity due to amplitude noise ($2\times 10^{-6}$) and phase mis-alignment ($2\times 10^{-5}$) is significantly lower than typical fault-tolerance thresholds such as those of the surface\cite{Fowler_2012}.
These results demonstrate that the amplitude and phase stability of QuEL-1 SE enables reliable long-duration quantum operations, thus highlighting its utility as a scalable control platform for superconducting and other qubit modalities.
\end{abstract}

\maketitle

\section{
\label{sec:intro}
Introduction
}

Coherent qubit control via microwaves is fundamental to the physical implementation of quantum information processing~\cite{bardin2021microwaves}.
In practice, microwave-based qubit control enables state initialization, gate operations, or readout across multiple hardware platforms including superconducting qubits~\cite{krantz2019quantum}, semiconductor spin qubits~\cite{burkard2023semiconductor}, 
nitrogen-vacancy centers~\cite{PhysRevApplied.23.034052}, and trapped atomic ions~\cite{bruzewicz2019trapped}.
In these systems, achieving high-fidelity qubit operations requires precise control of the frequency, amplitude, and phase of microwave signals.
Instabilities in these parameters, including drift and noise, directly degrade qubit operation fidelity.

The hardware that synthesizes and delivers microwave signals for qubit manipulation is commonly referred to as a qubit controller.
Such controllers must meet critical requirements that vary with qubit modality~\cite{bardin2021microwaves}. These
include wide frequency coverage with tunability, sufficient output power with fine amplitude resolution, low phase noise, and minimal interchannel crosstalk.
Satisfying these criteria is essential for performing accurate high-fidelity quantum operations across diverse qubit platforms.
Another important requirement is the ability to generate arbitrary waveforms that enable flexible pulse shaping to meet the demands of hardware platforms~\cite{vandersypen2004nmr, werschnik2007quantum, motzoi2009simple, chen2016measuring, hyyppa2024reducing, yi2024robust, matsuda2025selective, wang2025suppressing, zarantonello2019robust, weber2024robust}.
As the scale of quantum processors increases, qubit controllers must meet the requirements of system scalability and synchronization.
For example, each qubit must typically contain a dedicated microwave control channel in superconducting qubit systems ~\cite{krinner2019engineering, krantz2019quantum}.
Therefore, a microwave control system must provide a large number of output channels while maintaining precise synchronization across all the channels.

Controllers for various qubits have been developed in recent years.
Controllers for superconducting qubits include QICK~\cite{stefanazzi2022qick,ding2024experimental},
QubiC~\cite{xu2021qubic}, ICARUS-Q~\cite{park2022icarus}, and M2CS~\cite{zhang2024m2cs}, in addition to industrial systems such as Zurich Instruments QCCS~\cite{zurich2024qccs} and Keysight QCS~\cite{keysight2025qcs}.
Similarly, controllers for atomic qubits have been demonstrated~\cite{kulik2022latest, irtija2023design, maetani2024application}.

We have recently developed the QuBE~\cite{sumida2024qube, negoro2024experimental} scalable qubit controller for superconducting qubits. It features wideband outputs exceeding 1~GHz and supports unit-to-unit synchronization to enable system-level scaling.
The wide bandwidth of QuBE is particularly advantageous for the experiments that require access to broad transition spectra, including frequency-multiplexed qubit control~\cite{van2019impact, van2020scalable, ohira2024optimizing} and readout~\cite{PhysRevApplied.10.034040}, all-microwave unconditional qubit reset~\cite{magnard2018fast}, qudit operations~\cite{blok2021quantum, luo2023experimental}, and all-microwave gates between far-detuned qubits~\cite{krinner2020demonstration}.

These works have emphasized the high sampling rate of digital-to-analog converters (DACs) and analog-to-digital converters (ADCs), low spurious-free dynamic range, and low phase noise~\cite{sumida2024qube, negoro2024experimental}. 
However, negligible attention has been paid to the long-term stability of generated microwave signals. 
As the number of qubits increases and quantum error correction becomes feasible, which enable long-duration quantum operations, the stability of the microwave amplitude and relative phase becomes critical for achieving high-fidelity quantum gates.

For example, the QICK~\cite{ding2024experimental} control platform employs direct digital synthesis with phase reset at each experimental repetition to deterministically define the initial phase of all drives. This functionality ensures reproducible phase relationships within individual experiments. In contrast, the problem considered here concerns long-timescale operation, where phase coherence must persist over durations relevant to repeated error-correction cycles and system-level operation.
Similarly, in M2CS~\cite{zhang2024m2cs}, the amplitude stability is reported only for a single output signal. 
In addition, the sources and mitigation strategies for long-term variations in the microwave amplitude and phase have not been discussed in detail.

In this study, we focus on the temperature variations in analog devices, such as phase-locked loops (PLLs), amplifiers, and mixers, which are the primary sources of long-term variations in microwave signals. 
Our new qubit controller, hereafter referred to as QuEL-1~SE, incorporates a temperature stabilization system in which each analog device is equipped with an individual temperature-control loop. 
Each loop consists of a heater and thermistor for temperature sensing, along with a digital feedback controller. This maintains the device at a constant target temperature independent of ambient fluctuations. 

To evaluate the effectiveness of QuEL-1~SE, we simultaneously measure 15 microwave output signals using a common ADC over 24 h and analyze their amplitude and phase stability. Across the channels, the normalized amplitude exhibits standard deviations of 0.09\%--0.22\% (mean: 0.15\%), and the phase deviations are 0.35$^\circ$--0.44$^\circ$ (mean: 0.39$^\circ$). The gate infidelity caused by these microwave fluctuations is significantly lower than typical fault-tolerance thresholds.

The remainder of this paper is organized as follows:
The hardware architecture of the QuEL-1~SE system is described in Sec. ~\ref{sec:quel1se}.
The details of the temperature stabilization and active thermal management implementation are provided in Sec.~\ref{sec:temp_stab}.
The experimental setup and results are presented in Sec.~\ref{sec:exp_res}.
The discussion and conclusions are presented in Sec.~\ref{sec:discussion}.

\section{
\label{sec:quel1se}
Hardware Architecture of the QuEL-1~SE Microwave Controller
}

\begin{figure*}
    \centering
    \includegraphics[width=17cm]{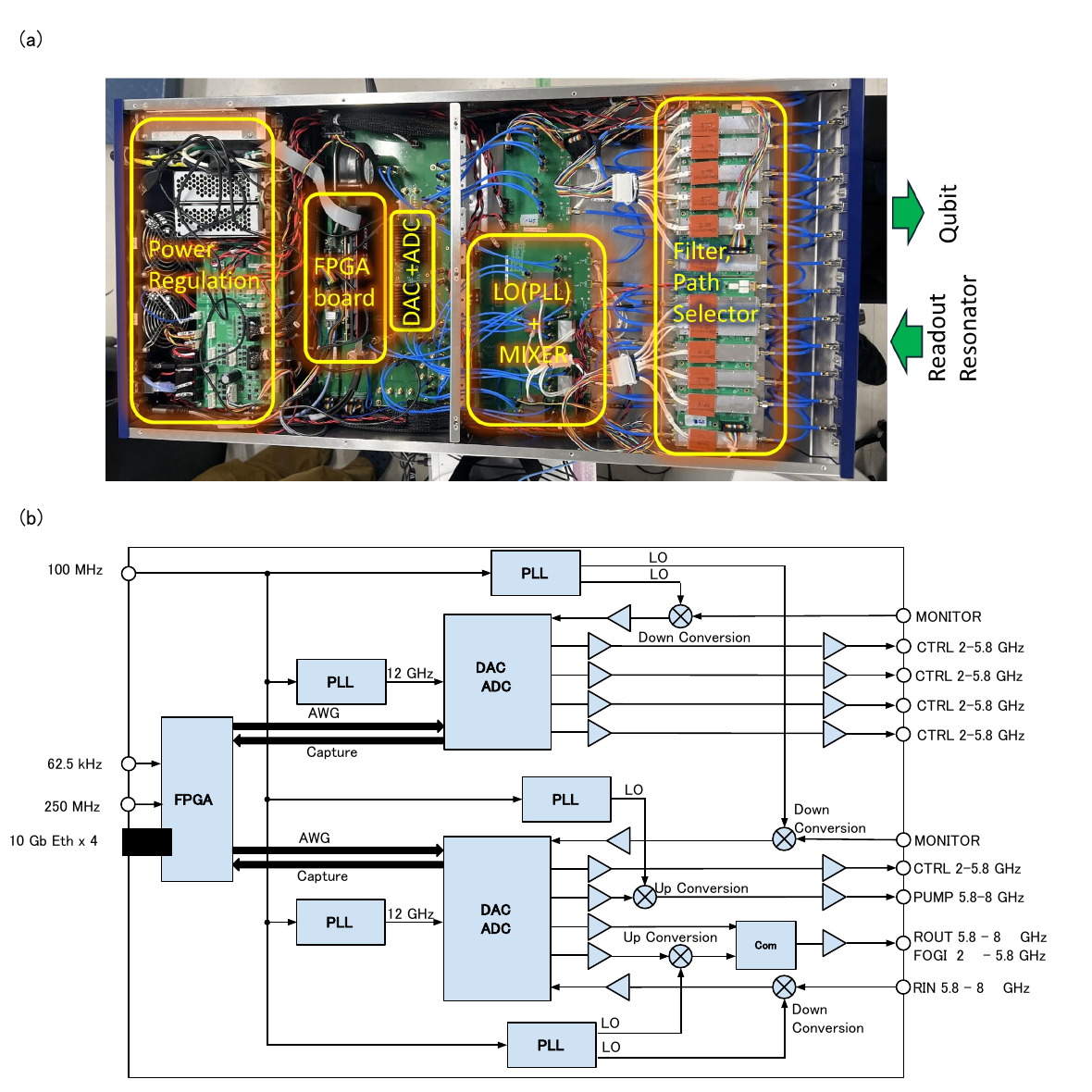}
    \caption{\label{fig:quel-1se}
   (a) Photograph of interior of QuEL-1~SE unit. 
   The unit is designed as a 19-inch rack-mount chassis with a height of 3U. 
   The height is perpendicular to the plane of the paper.
   Twelve brown rectangles visible in the "Filter, Path Selector" section are heaters used for temperature stabilization of the RF amplifiers. 
   Note that additional heaters for the PLL ICs and mixers are present but not visible. 
   Fans are placed to the left of the "Power Regulation" section.
   (b) Simplified block diagram of QuEL-1~SE unit. 
   Two AD9082 devices (labeled "DAC ADC") are used asymmetrically. The upper device handles the four \texttt{CTRL} ports without upconversion, whereas the lower device is used for \texttt{ROUT} and \texttt{PUMP} ports that require upconversion. 
   The QuEL-1~SE unit is optimized for four-multiplex readout of superconducting qubits. 
   For this application, one \texttt{CTRL} port of the lower device serves as an auxiliary channel. }
\end{figure*}

Figures~\ref{fig:quel-1se}(a) and (b) show a photograph and functional block diagram of the QuEL-1~SE microwave controller. 
The controller is designed for the generation, acquisition, and monitoring of the microwave signals required for manipulating superconducting qubits.
For clarity, the block diagram shows the essential elements of the architecture while omitting auxiliary components such as clock distribution integrated circuits (ICs), control buses, and per-port analog filtering (e.g., bandpass filters). 
The key performance specifications of QuEL-1~SE are summarized in Table~\ref{tab:quel1se-specs}. 
A defining feature of QuEL-1~SE is its wideband output range of 2~GHz to 8~GHz.

\subsection{Clocking and Frequency Reference Infrastructure}

Synchronization across multiple QuEL-1~SE units is achieved by sharing the following three clock signals:
(1) A 100~MHz reference clock distributed to all Texas Instruments LMX2594 PLL synthesizers.
(2) A 250~MHz clock used internally by a field-programmable gate array (FPGA).
(3) A 62.5~kHz clock used for synchronization with the global time counter.

\begin{figure*}
    \centering
    \includegraphics[width=17.0cm]{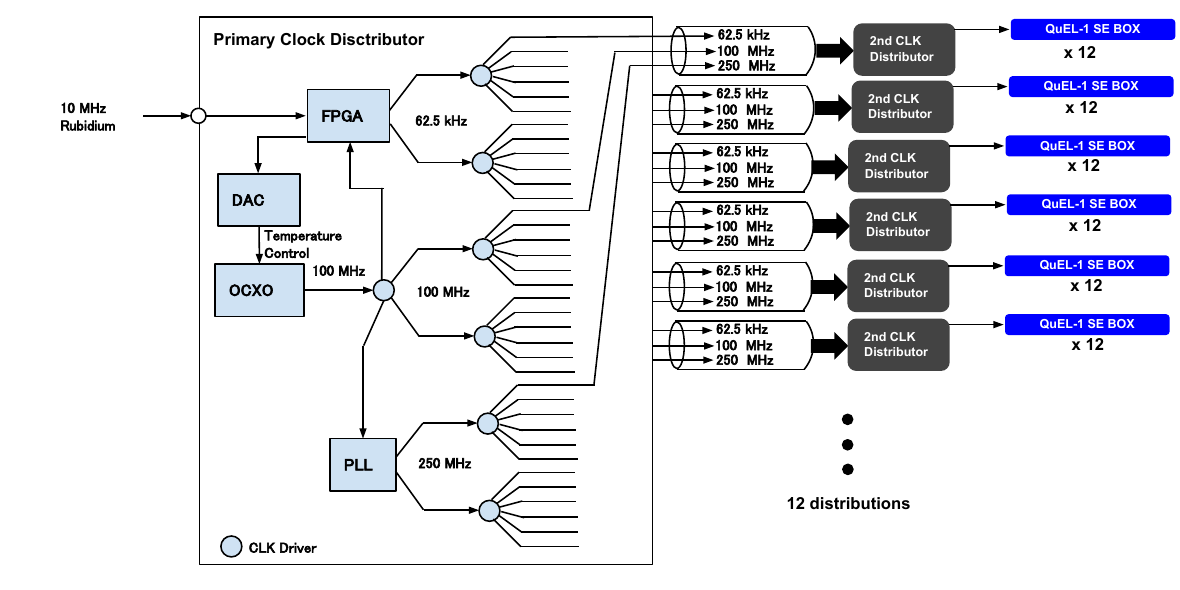}
    \caption{\label{fig:quel-1se-clocking}
Clock distribution network of the QuEL-1~SE system. 
To synchronize all derived clocks with the Rubidium oscillator, an oven-controlled crystal oscillator (OCXO) is employed in the primary clock distributor 
so that its frequency is stabilized by the temperature control of the OCXO. 
A compensator for fine frequency adjustment is implemented in the FPGA. 
    }
\end{figure*}

For this purpose, we develop a clock distributor that provides three types of clock signals to twelve channels, resulting in thirty-six output ports. 
Two cascaded distributors are implemented: a primary distributor, which uses a 10~MHz rubidium oscillator as its reference, and a secondary distributor, which derives its input from one set of the three distributed clocks.
This architecture enables the scalable expansion of the QuEL-1~SE systems.
Figure~\ref{fig:quel-1se-clocking} illustrates the overall clock distribution network and its connection to the QuEL-1~SE units.

\begin{table}[t]
    \caption{Key performance specifications of QuEL-1~SE.}
    \begin{tabular}{l l}
        \hline
        \textbf{Parameter} & \textbf{Specification} \\
        \hline
        Bandwidth (Control Port) & 2~GHz to 5.8~GHz \\
        Bandwidth (Readout/Input Port) & 5.8~GHz to 8~GHz \\
        Output Power & $\geq$ 0~dBm \\
        Channel-to-Channel Isolation & $-60$~dBc \\
        Noise Density & $-148$~dBm/Hz \\
        Phase Noise & $-100$~dBc/Hz @ 10~kHz offset \\
        \hline
    \end{tabular}
    \label{tab:quel1se-specs}
\end{table}

\subsection{Roles and Synchronization of LMX2594}

The system incorporates six LMX2594 chips with distinct roles.
Two of them generate 12~GHz reference signals for the DAC and ADC sampling clocks of the two AD9082 devices (Analog Devices, Inc.).  
The remaining four synthesize local oscillator (LO) signals for frequency upconversion and downconversion paths. 
The LO frequencies are selected according to application requirements.

\subsection{Data Converters and JESD204C Interface}

The QuEL-1~SE includes two AD9082chips. 
Each contains (1) 16-bit DACs that are capable of direct digital radio frequency (RF) synthesis up to 6~GHz and (2) 12-bit ADCs with input bandwidths of up to 3~GHz.
Each AD9082 chip communicates with the central FPGA via the JESD204C interface. 
The FPGA implements arbitrary waveform generation (AWG) and data capture to support quantum control and measurement.
The AWG operates at an effective sampling rate of 500~megasamples per second (MSPS). 

AD9082 features a two-stage upconversion and downconversion architecture. 
For the DAC, the first stage is a fine digital upconverter that upsamples the signal to 2~gigasamples per second (GSPS) using 4$\times$ interpolation and a numerically controlled oscillator. 
The second stage is a coarse digital upconverter that further upsamples the signal to 12~GSPS. 
QuEL-1~SE can simultaneously assign up to three AWGs, thereby allowing for the simultaneous generation of three independent 500~MHz bands within the 2~GHz range.
The process is reversed for the ADC. The signal is first downconverted by a coarse and then by a fine digital downconverter. 
This reduces the sampling rate from 6~GSPS to 500~MSPS.

\subsection{Signal Ports and Conversion Chains}

As summarized in Table~\ref{tab:port_info}, QuEL-1~SE features ten output ports and three input ports, which are categorized as follows:

\begin{table*}[t]
\caption{\label{tab:port_info}
Summary of input/output ports of QuEL-1~SE.
}
\begin{ruledtabular}
\begin{tabular}{ccccc}
\# & Name & Direction & Frequency Range & Description \\ \hline
0  & RIN         & In & 5.8--8.0~GHz  & Qubit response input \\
1  & ROUT / FOGI & Out & 5.8--8.0~GHz / 2.0--5.8~GHz & Readout and f0--g1 reset drive ($\ket{f,0} \leftrightarrow \ket{g,1}$) \\
2  & PUMP            & Out & 5.8--8.0~GHz & Pump signal output \\
3  & CTRL (extra)    & Out & 2.0--5.8~GHz & Auxiliary control \\
4  & MONITOR-in-0    & In & 2.0--8.0~GHz & Monitor input for Ports~1--3. \\
5  & MONITOR-out-0   & Out & 2.0--8.0~GHz & Monitor output for Ports~1--3. \\
6  & CTRL-0          & Out & 2.0--5.8~GHz & Control signal \\
7  & CTRL-1          & Out & 2.0--5.8~GHz & Control signal \\
8  & CTRL-2          & Out & 2.0--5.8~GHz & Control signal \\
9  & CTRL-3          & Out & 2.0--5.8~GHz & Control signal \\
10 & MONITOR-in-1    & In  & 2.0--5.8~GHz & Monitor input for Ports~6--9 \\
11 & MONITOR-out-1   & Out & 2.0--5.8~GHz & Monitor output for Ports~6--9 \\
\end{tabular}
\end{ruledtabular}
\end{table*}

\paragraph{CTRL Ports (Direct Outputs):}
Five output ports (CTRL) ranging from 2~GHz to 5.8~GHz are directly driven by the AD9082 DACs. 
These signals lie within the Nyquist zone and do not require upconversion. 
They are typically used to deliver quantum control pulses.

\paragraph{ROUT and PUMP Ports (Upconverted Outputs):}
The ROUT and PUMP ports (both cover a frequency range of 5.8--8~GHz) require upconversion because their target frequencies exceed the native output range of the DAC. 
The baseband waveform from the AD9082 chip is mixed with an LO synthesized from an LMX2594 chip to obtain the desired band. 
The ROUT port typically drives the readout resonator. In addition, can provide the f0--g1 reset drive (FOGI), which excites the $\ket{f,0} \rightarrow \ket{g,1}$ qubit-resonator transition. The subsequent readout resonator decay resets the system to $\ket{g,0}$. 
This protocol enables fast unconditional qubit initialization~\cite{magnard2018fast}.

\paragraph{RIN Port (Downconverted Input):}
The RIN port receives the signals reflected from the readout resonator. 
As the ADC bandwidth of the AD9082 is limited to 3~GHz, incoming signals in the 5.8--8.0~GHz range are first downconverted using a mixer. 
Note that the ROUT and RIN paths share a common LO generated by a single LMX2594 chip, thus ensuring a synchronized transmit--receive alignment.

\paragraph{MONITOR Port:}
The Input Monitor ports allow for the examination of the waveforms generated by multiple QuEL-1~SE units.
As the frequency range of 2--8~GHz exceeds the ADC bandwidth (3~GHz), the signals from the Input Monitor ports are downconverted prior to digitization through the ADC chain.
The Monitor Output ports provide diagnostic access to internal analog signals.
The combined use of the Input and Output Monitor ports enables concurrent diagnostics during qubit-control operations.

\section{
Temperature Stabilization and Active Thermal Management
}\label{sec:temp_stab}

\begin{figure}[t]
    \includegraphics[width=9.0cm]{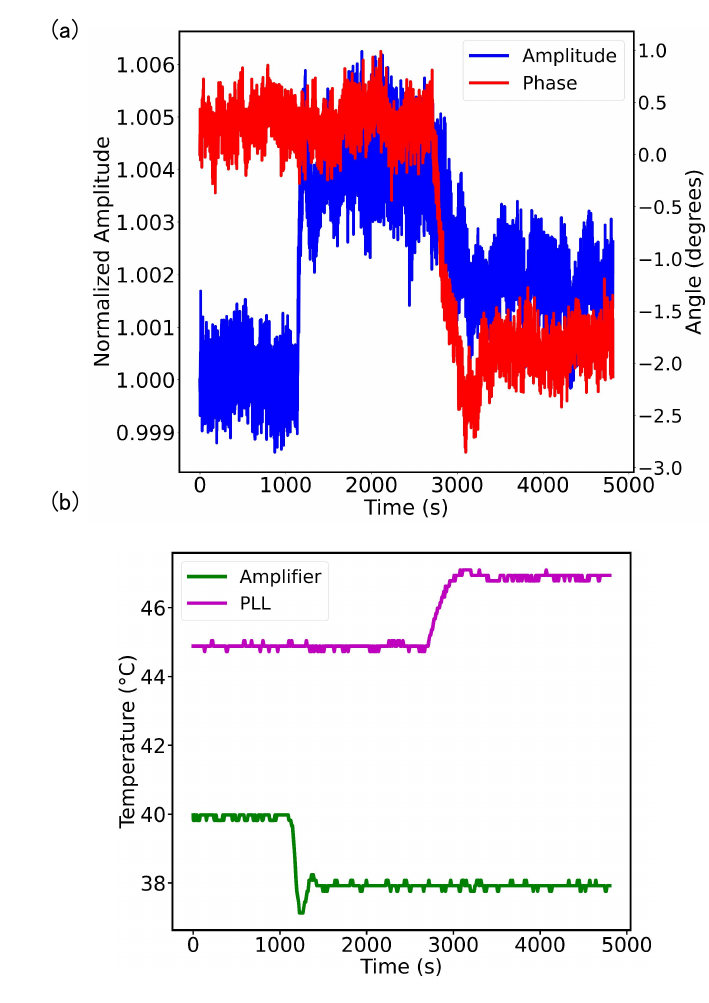}
    \caption{
    Step response of the output microwave to a temperature change. 
    (a) Normalized amplitude and phase. 
    (b) Temperatures around an amplifier and PLL IC. 
    The abrupt changes in device temperatures are induced by individual heaters placed close to the devices. 
    The temperatures are measured by thermistors placed close to the devices. 
    As the thermistors are mounted on the printed circuit board (PCB), these measurements do not directly represent the die temperatures.
    }
    \label{fig:amp_temp}
\end{figure}

As shown in Fig.~\ref{fig:amp_temp}, the changes in the temperature around the RF amplifier and PLL IC (LMX2594) result in a clear step response in the amplitude and phase of the output microwave, thus highlighting its sensitivity to thermal fluctuations.
Our experiments suggest that temperature variations in the amplifier affect the amplitude stability, whereas those in the PLLs influence the amplitude and phase stability.

\begin{figure}[t]
  \includegraphics[width=9.4cm]{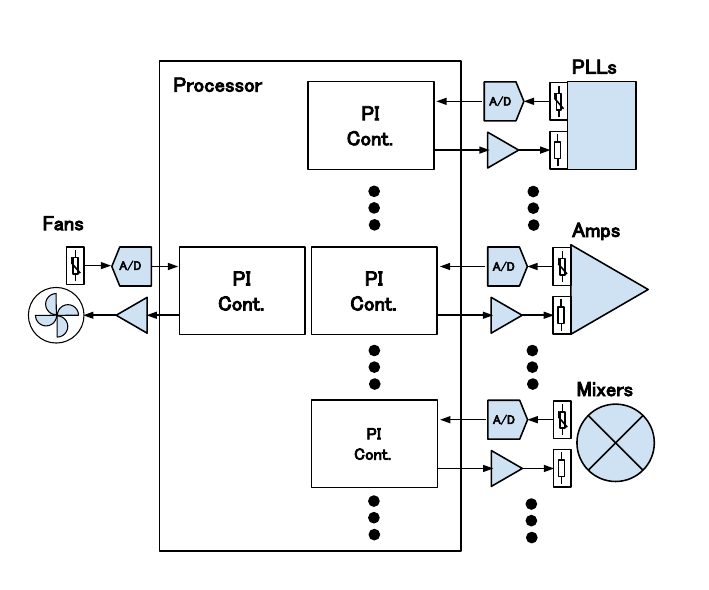}
  \caption{
  Active thermal management system of the QuEL-1~SE. 
  All measured temperatures are digitized using ADCs (Texas Instruments AD7490). 
  The processor compares each digitized temperature with its reference value and applies a proportional–-integral compensator to the difference to determine the PWM duty, which controls the power of the heaters and fans. 
  The heaters and fans are driven by individual MOSFET bridge circuits using PWM gate pulses.
  }
  \label{fig:quel-1se-thermal}
\end{figure}

The QuEL-1~SE controller implements a two-tiered temperature regulation strategy that combines global and local feedback control. 
This is a major update of our previous work~\cite{sumida2024qube, negoro2024experimental}.
At the system level, cooling fans installed on the enclosure provide active airflow. 
The fan speed is regulated using feedback from a thermistor placed at a representative location within the system. 
Although the fan is regulated based on a single temperature point, its airflow influences the temperature distribution across the enclosure and helps suppress large thermal drifts.

In addition to this global control, localized temperature regulation is implemented around the temperature-sensitive components, including low-noise amplifiers, LMX2594 chips, and RF mixers. 
Each of these critical components is equipped with a thermistor and heater. 
Independent feedback loops maintain each component at a constant temperature, thereby reducing the impact of thermal fluctuations on the phase and amplitude stability.

The temperature of each thermistor is digitized and used as an input to a proportional--integral compensator, which is individually implemented in a processor. 
The result of the compensator is used as the input of a pulse width modulation (PWM) amplifier to drive an actuator that is either a fan or a heater. 
The block diagram of the control system is shown in Fig.~\ref{fig:quel-1se-thermal}
This hybrid approach minimizes system-wide and component-level thermal variations, thereby contributing to the overall performance and long-term reliability of the controller.

\begin{figure}
  \includegraphics[width=0.48\textwidth]{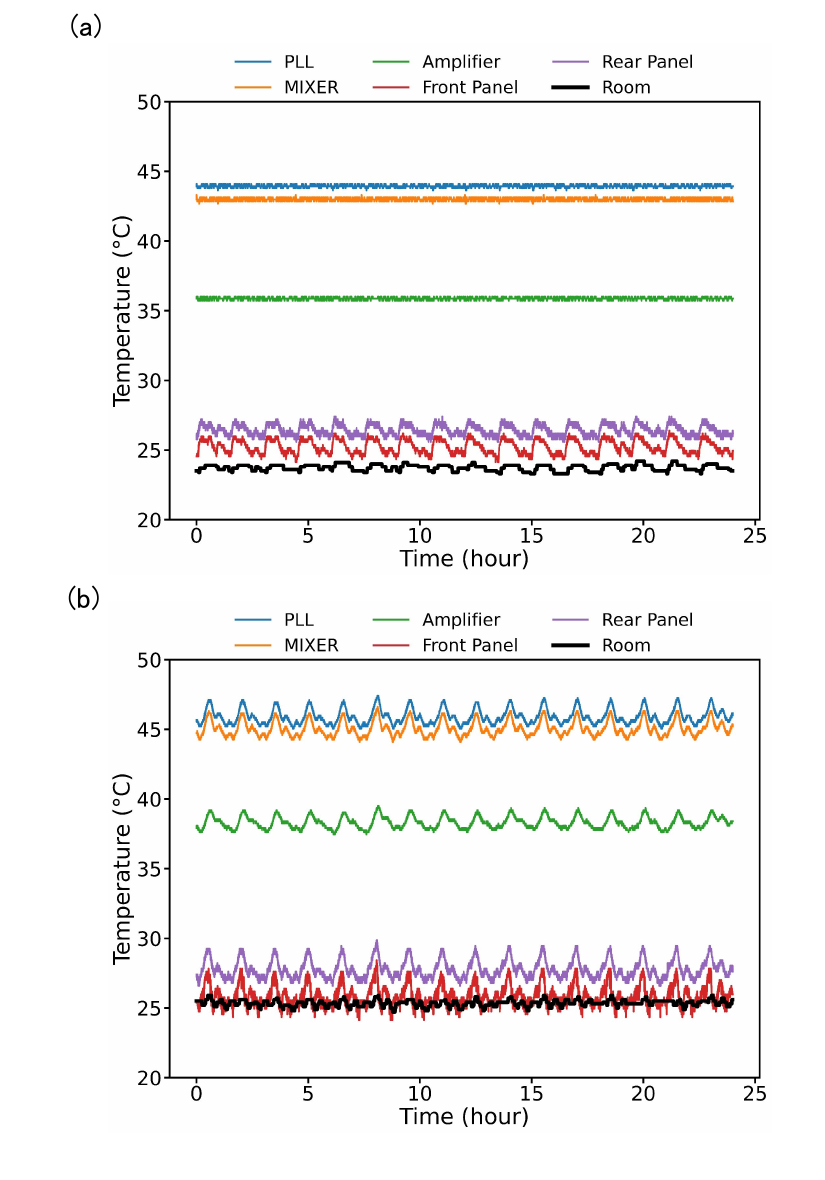}
  \caption{
  Component temperatures (PLL IC, amplifier, and mixer), air temperatures at the panels and room (a) with and (b) without active thermal control. 
  The room temperature is shown for reference. 
  When using typical air conditioners, the room temperature fluctuates within a range of 1--2~\textcelsius{} with a period of several tens of minutes.
  }
  \label{fig:temperatures}
\end{figure}

The temperatures of the components with and without active thermal control are shown in Fig.~\ref{fig:temperatures}.
The components clearly show larger variations without active thermal control, and similar trends are observed for the room temperature.
When typical air conditioners are used, the variation in the room temperature can be approximately characterized by an amplitude of 1--2~\textcelsius{} 
and a period of several tens of minutes. 
The fluctuation pattern depends on the room environment and outdoor temperature. 
For practical use, the QuEL-1~SE units are assumed to be installed in rooms with standard air-conditioning systems. 
They do not require precise air conditioning.

\section{
\label{sec:exp_res}
Long-term Simultaneous Measurement of 15 Microwave Outputs  
}

Reliable microwave sources with excellent amplitude and phase stability are essential for high-fidelity qubit control. 
We quantify the stability of the generated microwaves across 15 channels over 24 h.

\subsection{
Experimental Configuration 
}

\begin{figure}
  \includegraphics[width=8.0cm]{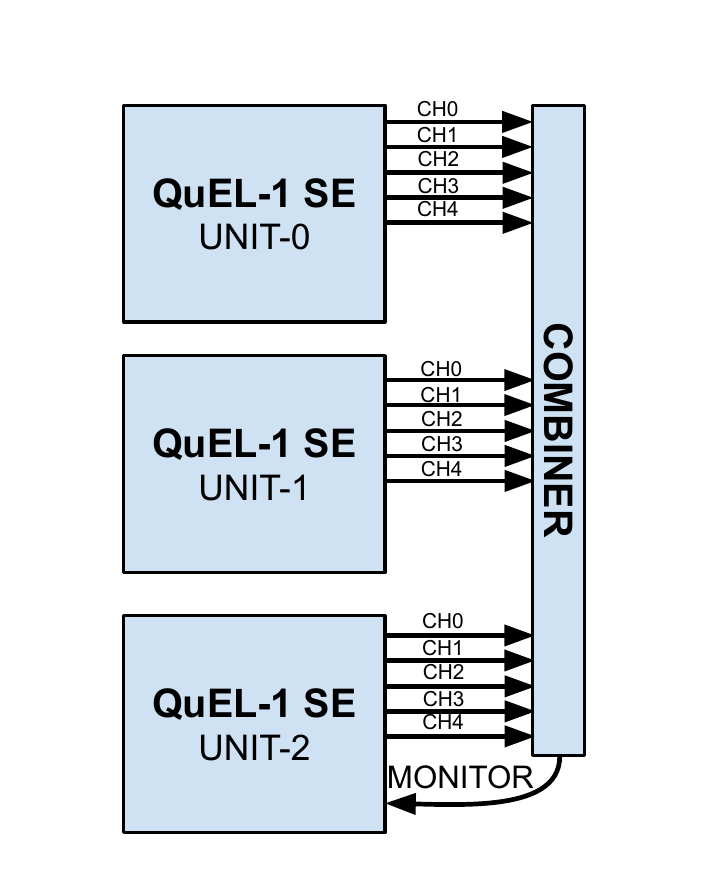}
  \caption{
  Experimental setup for long-term simultaneous measurement of 15 microwave outputs.
  Five \texttt{CTRL} channels from each of the three QuEL-1~SE units are combined and connected to the Monitor Input port of one of the units. 
  The units are labeled as "UNIT-0,1,2" and the channels as "CH0,1,2,3,4."}
  \label{fig:experimental_setup}
\end{figure}

\begin{figure}
  \centering
  \includegraphics[width=9.0cm]{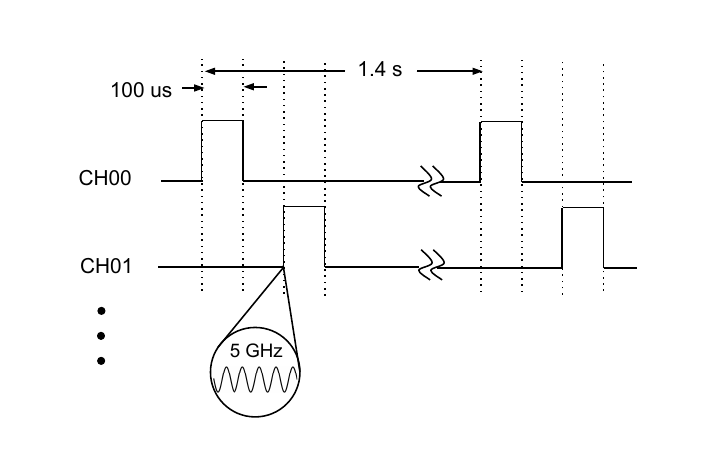}
  \caption{
  Time-domain relationship among the 15 microwave outputs. 
  Each of the 15 channels generates \SI{100}{\micro\second} pulses at different timings. 
  The pulse-to-pulse interval is set as \SI{100}{\micro\second}. 
  The ADC captures the combined waveform composed of 15 \SI{100}{\micro\second} pulses.
  }  
  \label{fig:waveform}
\end{figure}

As shown in Fig.~\ref{fig:experimental_setup}, three QuEL-1~SE units are configured, where each unit provides five \texttt{CTRL} output 
channels, i.e., a total of 15 channels (indexed 0--14). 
The 15 outputs are routed to a single ADC in one unit. 
The acquisition is time multiplexed. Each channel emits a \SI{100}{\micro\second} pulse at \SI{5}{\giga\hertz} approximately 
every $\sim$\SI{1.3}{\second}, and the ADC records all channels in sequence. 
Figure~\ref{fig:waveform} shows the time-domain relationship between  
the 15 microwave outputs. 

The output power of all the emitted microwaves is adjusted to exceed 0~dBm at the ports of the QuEL-1~SE units.
For each captured pulse, we computed the complex envelope and extract
(i) the \emph{dimensionless} mean amplitude (normalized per channel) and (ii) mean phase in degrees. For each channel, the per-pulse mean amplitude is normalized to enable direct comparison across the channels. 
The phase is reported in degrees.
We use the per-pulse time series to derive two scalar stability metrics per channel, i.e., the peak-to-peak variation and standard deviation, which are  separately computed for the amplitude and phase.

There are two reasons for selecting only \texttt{CTRL} channels and not using the \texttt{ROUT} channels.
First, an \texttt{ROUT} channel is always used with an \texttt{RIN} channel that shares a common PLL IC. Thus, measuring \texttt{ROUT} with \texttt{RIN} underestimates the actual microwave fluctuations due to correlated PLL noise, leading to an unfair comparison with \texttt{CTRL} channels.
Second, although the signals from the \texttt{ROUT} channels pass through the upconversion stage, which is not included in the \texttt{CTRL} channels, the temperature variations in the mixer and LO (LMX2594) are already considered in the \texttt{CTRL} channels in the downconversion stage before digitization as shown in Fig.~\ref{fig:quel-1se}.

\subsection{Results}
\begin{figure*}
    \includegraphics[width=17cm]{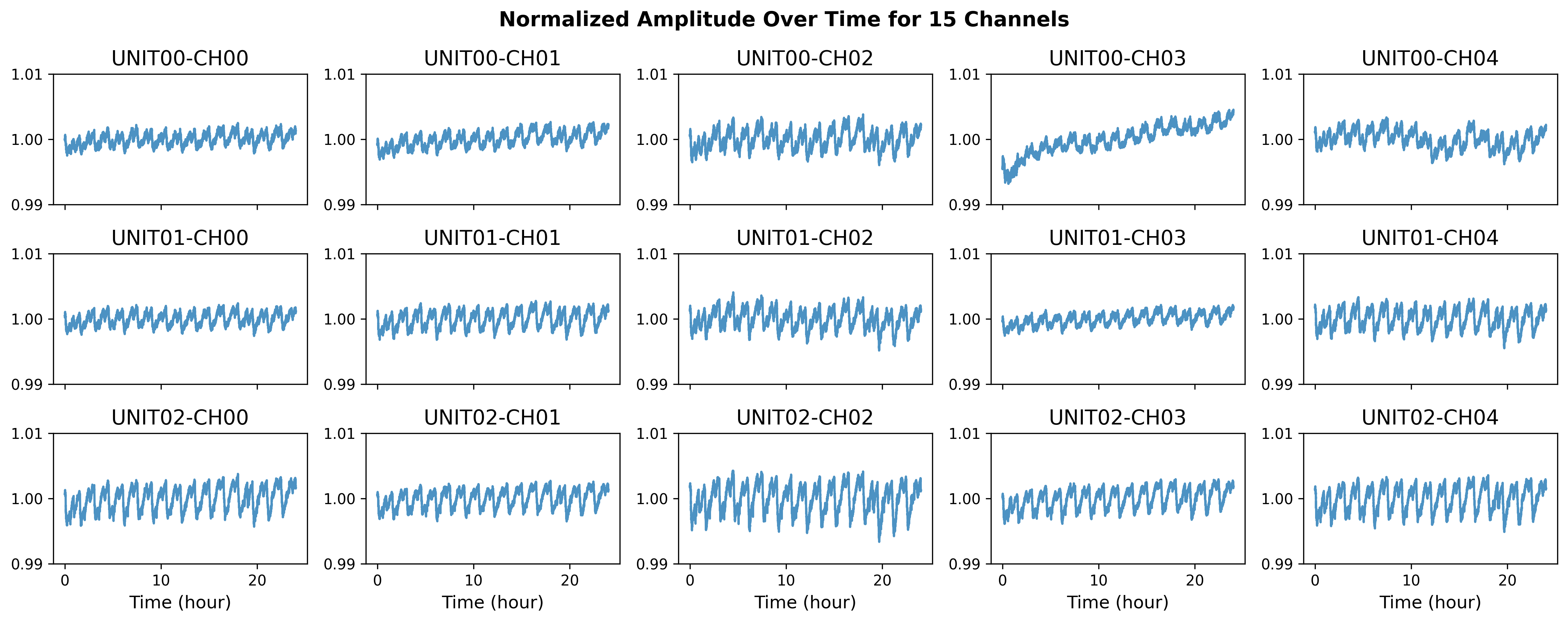}
    \caption{
    Normalized amplitude versus time for channels 0--14. 
    Each trace corresponds to one channel. 
    Periodic patterns reflecting the room-temperature fluctuation are observed even with the stabilization of the device temperatures. 
    However, the amplitude variations are considerably smaller than those in the case without stabilization, as shown in Fig.~\ref{fig:amplitude_and_angle_comparison} (a)}
    \label{fig:amp}
\end{figure*}

\begin{figure*}
    \includegraphics[width=17cm]{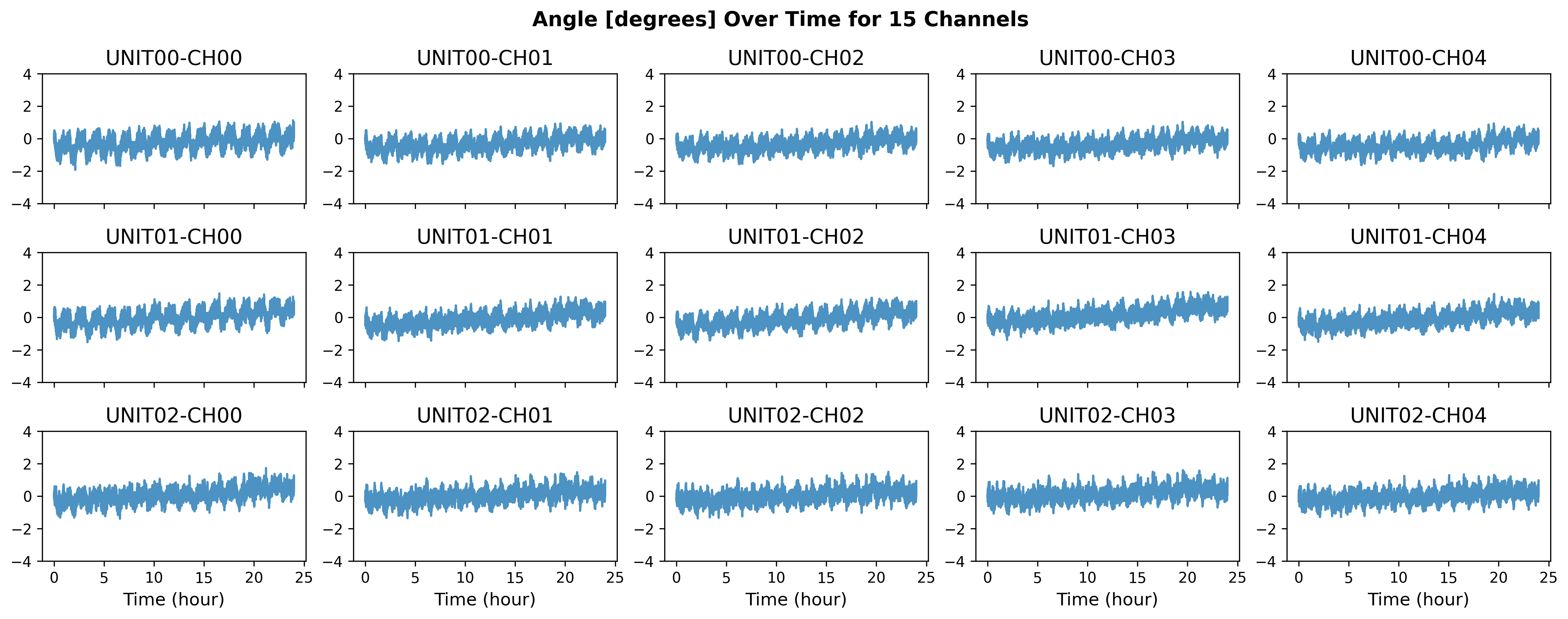}
    \caption{
    Phase (degrees) versus time for channels 0--14. 
    Each trace corresponds to one channel. The periodic patterns are observed although the variations are considerably smaller than those in the case without stabilization, as shown in Fig.~\ref{fig:amplitude_and_angle_comparison} (b)}
    \label{fig:phase}
\end{figure*}

Figure~\ref{fig:amp} shows the normalized amplitude versus time for all channels, and Fig.~\ref{fig:phase} shows the corresponding phase (degrees) versus time.
A statistical summary is provided in Table~\ref{tab:stats}. 
Across the channels, the standard deviation of the amplitude is \SIrange{0.09}{0.22}{\percent} (mean \SI{0.15}{\percent}) and that of the phase is \SIrange{0.35}{0.44}{\degree} (mean \SI{0.39}{\degree}).

\begin{figure}
  \centering
  \includegraphics[width=8cm]{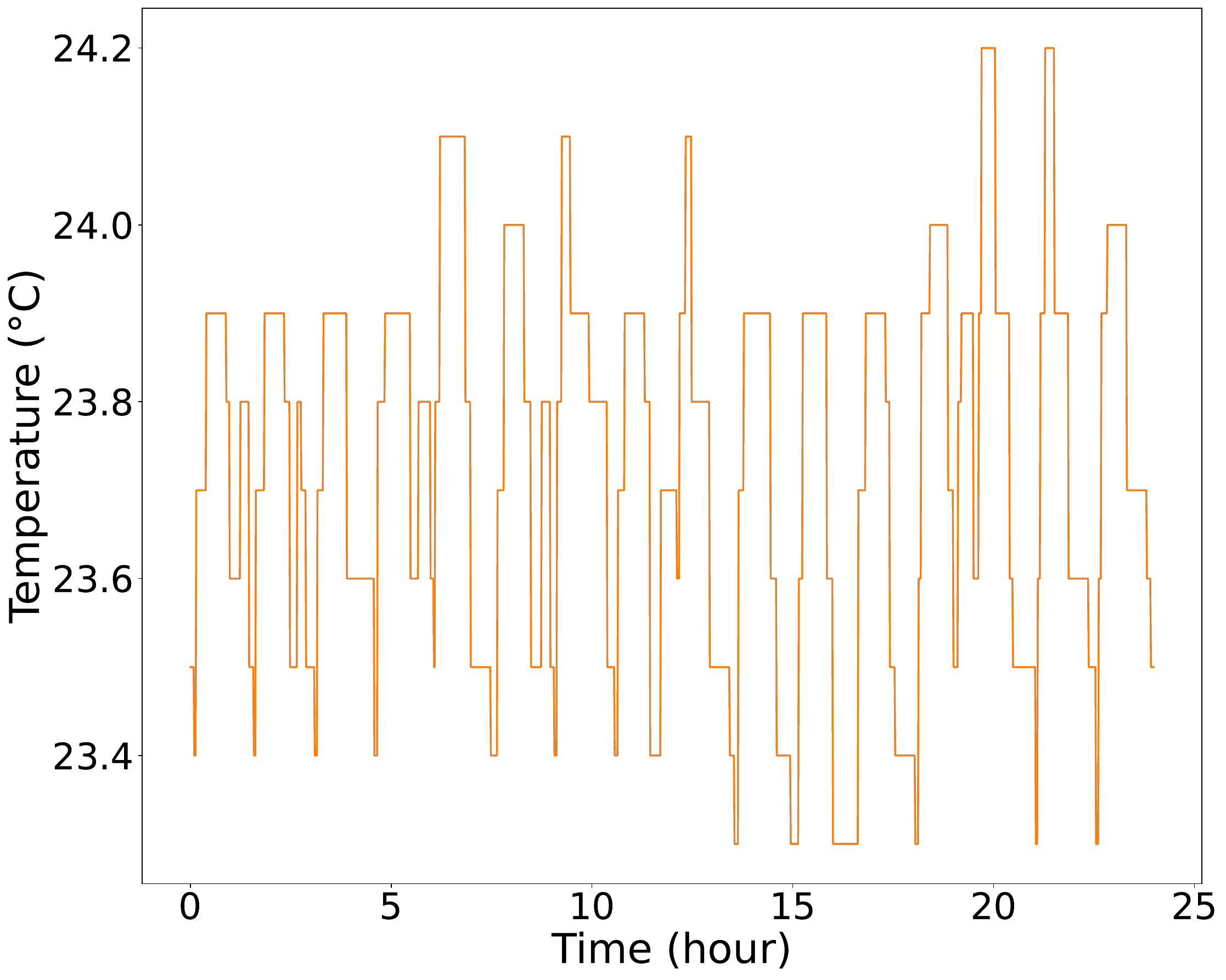}
  \caption{Room temperature measured over 24 h during the experiments. }
  \label{fig:room_temperature}
\end{figure}

There are still gradual variations in the amplitude and phase. 
The period of these variations coincides with that of 
the room temperature variations. 
Figure~\ref{fig:room_temperature} shows the room temperature measured for 24 h during the experiments.
These variations are extremely common when the room temperature is controlled by a normal air-conditioning system.
%

\begin{table*}[t]
\caption{
Stability statistics per channel. ``Amp'' denotes normalized amplitude (\%).
}
\label{tab:stats}
\begin{ruledtabular}
\begin{tabular}{
    c
    c
    S[table-format=1.2]
    S[table-format=1.3]
    S[table-format=1.1]
    S[table-format=1.2]
}
Unit & Channel &
{Amp Peak-to-Peak (\%)} &
{Amp Std.\ Dev.\ (\%)} &
{Phase Peak-to-Peak (deg)} &
{Phase Std.\ Dev.\ (deg)} \\
\hline
0 & 0  & 0.50 & 0.092 & 3.1 & 0.43 \\
0 & 1  & 0.58 & 0.110 & 2.5 & 0.37 \\
0 & 2  & 0.78 & 0.150 & 2.6 & 0.36 \\
0 & 3  & 1.10 & 0.220 & 2.7 & 0.35 \\
0 & 4  & 0.72 & 0.140 & 2.6 & 0.36 \\
1 & 0  & 0.50 & 0.098 & 3.0 & 0.44 \\
1 & 1  & 0.60 & 0.130 & 2.7 & 0.40 \\
1 & 2  & 0.89 & 0.150 & 2.8 & 0.42 \\
1 & 3  & 0.47 & 0.090 & 3.0 & 0.41 \\
1 & 4  & 0.78 & 0.150 & 3.0 & 0.40 \\
2 & 0 & 0.81 & 0.180 & 3.1 & 0.41 \\
2 & 1 & 0.62 & 0.140 & 2.8 & 0.37 \\
2 & 2 & 1.10 & 0.220 & 2.9 & 0.37 \\
2 & 3 & 0.69 & 0.150 & 2.8 & 0.37 \\
2 & 4 & 0.87 & 0.190 & 2.6 & 0.36 \\
\end{tabular}
\end{ruledtabular}
\end{table*}

\subsection{
Impact of Active Thermal Control
}

\begin{figure}
  \includegraphics[width=0.48\textwidth]{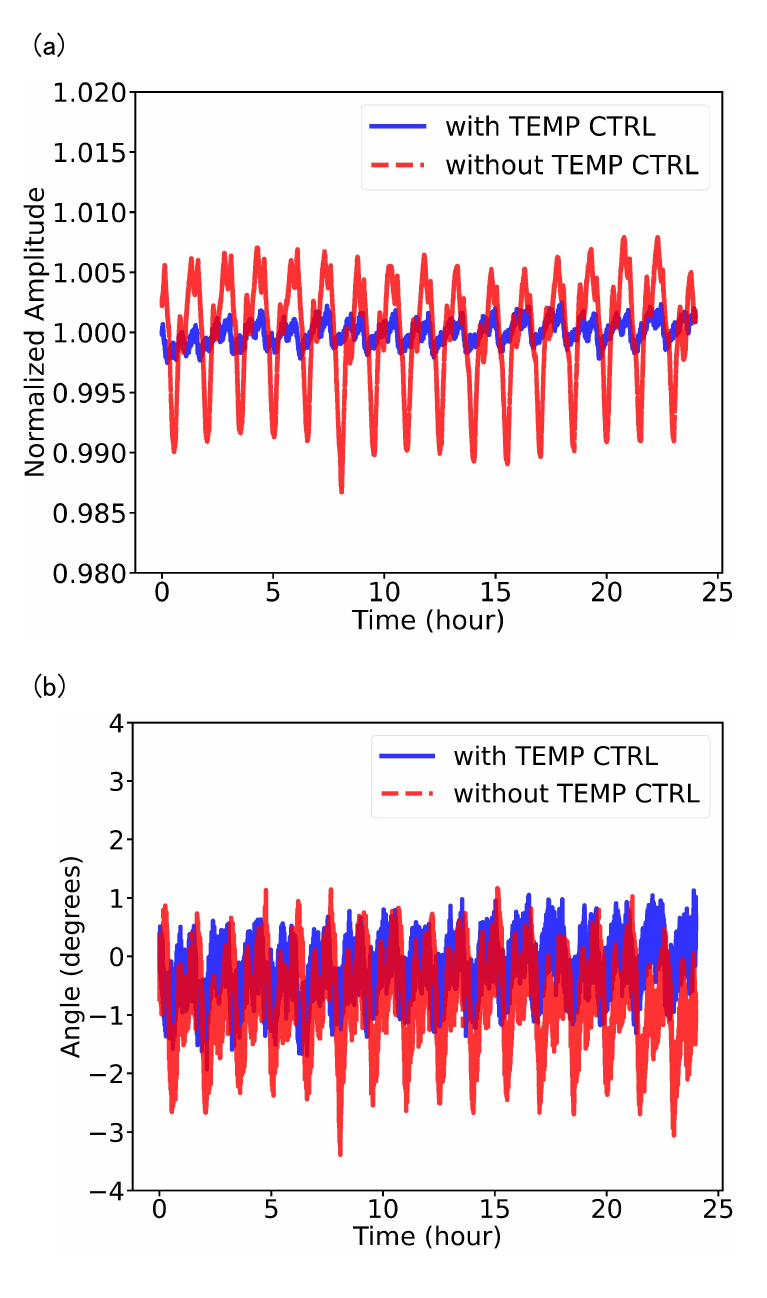}
  \caption{Amplitude and phase comparison (a) with and (b) without active thermal control. 
  Only one channel (UNIT00–CH00) is shown for simplicity. 
  }
  \label{fig:amplitude_and_angle_comparison}
\end{figure}

The measurements demonstrate subpercent amplitude stability across all channels and phase variations that are considerably lower than a degree. 
To assess the impact of active thermal control, we repeat the experiment for another 24 h with the thermal control \emph{disabled} (constant power applied to all actuators: fans and heaters). 
In this case, the standard deviation of the amplitude is \SIrange{0.35}{0.56}{\percent} (mean \SI{0.45}{\percent}). 
and that of the phase is \SIrange{0.70}{0.79}{\degree} (mean \SI{0.73}{\degree}). 
Thus, the amplitude and phase deviations are more than doubled without active thermal control.
Figure~\ref{fig:amplitude_and_angle_comparison} compares the amplitude and angle with and without active thermal control.

\section{
\label{sec:discussion}
Discussion and Conclusion
}

We quantify gate errors using the average gate fidelity between ideal unitary $U$ and implemented operation $V$, as follows:
\begin{equation}
F_{\mathrm{avg}}(U,V)
= \frac{|\mathrm{Tr}(U^\dagger V)|^2 + d}{d(d+1)},\qquad d=2.
\label{eq:favg_def}
\end{equation}
For a target $X/2$ gate, $U_{X/2}=\exp(-i\frac{\pi}{4}X)$, we consider two coherent error models that correspond to the observed stability metrics: (i) amplitude (over/under rotation) error and (ii) axis-phase misalignment error.

\emph{(i) Amplitude error.}
The amplitude (rotation angle) error of fractional amount $\epsilon$ is expressed as
\begin{equation}
    V_{\mathrm{amp}}(\epsilon)=\exp\!\left[-i\left(\frac{\pi}{4}\left(1+\epsilon\right)\right)X\right],
    \quad \delta\theta=\frac{\pi}{2}\epsilon.
\end{equation}
Using Equation~\eqref{eq:favg_def} with $U=U_{X/2}$ and $V=V_{\mathrm{amp}}(\epsilon)$ gives
\begin{equation}
    1-F_{\mathrm{avg}}^{(\mathrm{amp})}(\epsilon)
    = \frac{2}{3}\sin^2\!\left(\frac{\delta\theta}{2}\right)
    \simeq \frac{\delta\theta^2}{6},
\label{eq:f_amp_pi2}
\end{equation}
where the last expression is the small-angle approximation.  
With the measured standard deviation, $\epsilon=0.22\,\%=2.2\times10^{-3}$, $1-F_{\mathrm{avg}}^{(\mathrm{amp})}\simeq 2.0\times10^{-6}$, whereas without thermal control ($\epsilon=0.45\,\%$) it increases to $8.3\times10^{-6}$.

\emph{(ii) Axis-phase misalignment.}
A phase misalignment of $\phi$ (rotation by $\pi/2$ about the axis $\cos\phi\,X+\sin\phi\,Y$) is expressed as
\begin{equation}
V_{\mathrm{phase}}(\phi)=\exp\!\left[-i\frac{\pi}{4}\left(\cos\phi\,X+\sin\phi\,Y\right)\right].
\end{equation}
Equation~\eqref{eq:favg_def} yields
\begin{equation}
1-F_{\mathrm{avg}}^{(\mathrm{phase})}(\phi)
= \frac{2}{3}\left[1-\cos^4\!\left(\frac{\phi}{2}\right)\right]
\simeq \frac{1}{3}\phi^2,
\label{eq:f_phase_pi2}
\end{equation}
where the last expression holds for small values of $\phi$.  
With the measured standard deviation, $\phi=0.44^{\circ}=7.67\times10^{-3}\,\mathrm{rad}$, $1-F_{\mathrm{avg}}^{(\mathrm{phase})} \simeq 2.0\times10^{-5}$, whereas without thermal control ($\phi=0.79^{\circ}$) it increases to $6.3\times10^{-5}$.

In summary, based on the observed amplitude and phase fluctuations, the estimated contribution to the $X/2$-gate infidelity due to amplitude noise ($\sim2\times10^{-6}$) and phase misalignment ($\sim2\times10^{-5}$) is small.
While these values are not experimentally verified through direct qubit measurements, they suggest that the achieved microwave stability would not be a limiting factor for single-qubit gate performance, even for future superconducting qubits with substantially improved coherence times.

Although the QuEL-1 SE system is originally developed for superconducting qubits, it can be extended to other qubit modalities that employ microwave control, such as trapped ions~\cite{miyoshi2024microwave}. Furthermore, it can be synchronized with control instruments in other electromagnetic domains, including optical and DC-voltage systems, to enable unified operation across diverse control frequencies~\cite{miyoshi2025toward}.

\begin{acknowledgments}
This work was supported by JST COI-NEXT (Grant No. JPMJPF2014), JST Moonshot R\&D (Grant No. JPMJMS226A), and MEXT Q-LEAP (Grant No. JPMXS0120319794). 
\end{acknowledgments}

\section*{Author Contributions}
\textbf{Yoshinori~Kurimoto:} Conceptualization (lead); Data curation (lead); Formal analysis (lead); Investigation (lead); Methodology (lead); Project administration (equal); Resources (equal); Software (equal); Validation(lead); Visualization (lead); Writing/Original Draft Preparation (equal); Writing/Review \& Editing (equal).
\textbf{Dongjun~Lee:} Formal analysis (supporting).
\textbf{Koichiro~Ban:} Resources (supporting).
\textbf{Shinichi~Morisaka:} Resources (equal).
\textbf{Toshi~Sumida:} Resources (supporting).
\textbf{Hidehisa~Shiomi:} Writing/Review \& Editing (supporting).
\textbf{Yosuke~Ito:} Resources (supporting).
\textbf{Yuuya~Sugita:} Resources (equal); Software (equal).
\textbf{Makoto~Negoro:} Funding Acquisition (equal), Writing/Review \& Editing (equal).
\textbf{Ryutaro~Ohira:} Conceptualization (supporting); Investigation (supporting); Methodology (supporting); Project administration (equal); Supervision (lead); Writing/Original Draft Preparation (equal); Writing/Review \& Editing (equal).
\textbf{Takefumi~Miyoshi:} Funding Acquisition (equal), Writing/Review \& Editing (equal).

\section*{data availability statement}
The data that support the findings of this study are available
from the corresponding author upon reasonable request.

\bibliographystyle{apsrev4-2}
\bibliography{main}

\end{document}